\documentstyle[12pt]{article}

\title{
Decay on several sorts of heterogeneous centers:
Special monodisperse approximation in the situation
of strong unsymmetry. 2. Numerical results for the total
monodisperse approximation}

\author{V.Kurasov}

\date{Victor.Kurasov@pobox.spbu.ru}

\begin{document}

\maketitle

\section{Preliminary remarks}

In \cite{Section1} we have investigated the process of nucleation in the
situation of the strong unsymmetry.  We have analysed the system of
condensation
equations and suggested three different approximations.

The first approximation is the total monodisperse approximation. It has
been already suggested in \cite{Multidecay} and is a rather natural one.
In this approximation the total number of the droplets on the
first type centers are regarded as  those  formed  at  the  initial
moment
of time. Then all these droplets have now one and same size which can
be easily calculated. It equals to $z$.

Certainly this approximation is suitable in the case of the strong unsymmetry.
Namely in this case it was used in \cite{Multidecay}. But this approximation
can be applied in some other cases.
This approximation can be used  to estimate
the errors of some other approximations.

It is clear that the total monodisperse approximation is more rough than
the special monodisperse approximation and the floating monodisperse
approximation.
Then we shall estimate the errors of the mentioned approximations by the
the error of the total monodisperse approximation.

In the special monodisperse approximation we have to
introduce the characteristic  size $\Delta z$ of the length of the
spectrum due to the supersaturation fall. This value is well described
in \cite{Multidecay}, \cite{Section1}. Then we have to imagine that the
influence of the droplets formed on the first type centers can be described
as the monodisperse peak with the number of droplets determined by the
special recipe.

To determine the number of droplets in the monodisperse spectrum we must
calculate the number of the first type heterogeneous centers which became
the centers of the droplets until $\Delta z / 4$. It can be done without
any influence of the second type heterogeneous centers taken into account.

The reason of concrete choice of the size $\Delta z / 4 $ is described
in \cite{Multidecay} in details. So, we needn't to explain it here.
We have only to note that this choice is equivalent to the  specific choice
of time, i.e. one has to calculate the number of the droplets on the first
type centers formed until the first quarter of the nucleation period. More
rigorously we have to speak here about the finish of the nucleation period
due to the fall of supersaturation.

In fact we can act without the special monodisperse approximation but
only
with the help of the floating monodisperse approximation. This approximation
is similar to the already described one but has one specific feature.
 In the floating
monodisperse approximation the influence of the droplets formed on the
first type centers at the "moment" $z$ can be presented as $z^3 N_1 (z/4)$,
where $N_1 (z/4) $
is the number of the droplets formed on the first type centers until $z/4$.
So, this approximation is formulated for all moments of time and can be
used in the arbitrary moment.

Certainly when $z$ is near $\Delta z$ this approximation coincides with
the special monodisperse approximation. But this approximation is more
simple and universal than the previous one. We shall use below the floating
monodisperse approximation instead of the special monodisperse one.

\section{Calculations for the total monodisperse approximation}

Now we shall turn to estimate errors of approximations. The errors of
substitutions of the subintegral functions by the rectangular form are
known and they are rather small. But the error of the
approximation itself has to be estimated.

The error of the number of droplets formed on the first type of heterogeneous
centers can be estimated in frame of the standard iteration method and
it is small. So, we need to estimate only the error in the number of the
droplets formed on the second type of heterogeneous centers.

It is absolutely clear that the worst situation occurs when there is no
essential
exhaustion of heterogeneous centers of the second type.

It seems that the monodisperse approximation will be the worst for the
pseudo homogeneous  situation, i.e. when the first type centers remain
practically
unexhausted. But as far as we haven't any direct proof of this property we
shall consider the situation with the arbitrary power of exhaustion.

As the result we can consider the system of the following form
$$
G = \int_0^z \exp(-G(x)) \theta_1(x) (z-x)^3 dx
$$
$$
\theta_1 = exp(-b \int_0^z \exp(-G(x)) dx)
$$
with a positive parameter $b$ and
have to estimate the error in
$$
N = \int_0^{\infty} \exp(-l G(x)) dx
$$
with some parameter $l$.

Parameter $l$ shows that we doesn't consider the influence  of the first
centers nucleation on itself but analyze the influence of the first centers
nucleation on another process with another parameters. This differs our
consideration from that made in \cite{Pavlov-uni}.

We shall solve this problem numerically  and compare our result with some
models. In the model of the total monodisperse approximation we get
$$
N_A = \int_0^{\infty} \exp(-l G_A(x)) dx
$$
where
$G_A$ is
$$
G_A = \frac{1}{b} (1 - \exp(-b D)) x^3
$$
and the constant $D$ is given by
$$
D = \int_0^{\infty} \exp(-x^4 /4) dx = 1.28
$$

We have tried the total monodisperse approximation
for $b$ from $0.2$ up to $5.2$
with the step $0.2$ and for $l$ from $0.2$ up to $5.2$ with a step $0.2$.
We calculate the relative error in $N$. The results are drawn in fig.1
for $N_A$.    Here  $r_1$  is  the  relative  error  of  the  total
monodisperse
approximation in the number of droplets.

We see that even for $N_A$ the relative error is small for all situations
with moderate $l$. For big  $l$ the error slightly increases. This corresponds
to the evident fact that when $l$ is big the nucleation on the second
sort centers is finished earlier than on the first sort centers. Here
the total monodisperse approximation isn't valid.

The growth of the error for big $l$ is rather slow but inevitably it will
lead to
 the big  value. Then
this approximation will give a wrong result even in the
order of the magnitude.

The calculations presented in  further sections show that the
 maximum of errors in the floating monodisperse approximation
lies near $l=0$. So, we have
to analyse the situation with small values of $l$. It was done in fig.2
for $N_A$.
We see that for $N_A$ this situation is even better than the previous
situation. It is rather natural because the  small  values of $l$ correspond
to more strong ierarchy.

Unfortunately
the situation for the floating monodisperse approximation is another.
We can not
find the maximum error
of the results for the monodisperse approximation. We see that this error
has the maximum at small $b$.
Then we  have to calculate the situation with $b=0$.
Here we have to solve the following equation
$$
G = \int_0^{\infty} \exp(-G(x)) (z-x)^3 dx
$$
and to compare
$$
N = \int_0^{\infty} \exp(-l G) dx
$$
with
$$
N_A = \int_0^{\infty}
\exp(-l D z^3)        dz
$$
and other approximate expressions.

The results of this calculation are interesting mainly for floating
monodisperse
approximation and will be presented in the next sections.

\pagebreak

\begin{picture}(350,350)
\put(0,0){\line(0,1){350}}
\put(0,0){\line(1,0){350}}
\put(350,0){\line(0,1){350}}
\put(0,350){\line(1,0){350}}
\put(152,235){.}
\put(158,233){.}
\put(164,230){.}
\put(170,228){.}
\put(176,226){.}
\put(182,224){.}
\put(188,222){.}
\put(194,221){.}
\put(200,219){.}
\put(206,218){.}
\put(212,216){.}
\put(218,215){.}
\put(224,214){.}
\put(230,213){.}
\put(236,212){.}
\put(242,211){.}
\put(248,210){.}
\put(254,209){.}
\put(260,208){.}
\put(266,208){.}
\put(272,207){.}
\put(278,206){.}
\put(284,206){.}
\put(290,205){.}
\put(296,205){.}
\put(302,204){.}
\put(148,236){.}
\put(154,234){.}
\put(160,231){.}
\put(166,229){.}
\put(172,227){.}
\put(178,225){.}
\put(184,223){.}
\put(190,221){.}
\put(196,219){.}
\put(202,217){.}
\put(208,216){.}
\put(214,214){.}
\put(220,213){.}
\put(226,212){.}
\put(232,211){.}
\put(238,210){.}
\put(244,209){.}
\put(250,208){.}
\put(256,207){.}
\put(262,206){.}
\put(268,205){.}
\put(274,205){.}
\put(280,204){.}
\put(286,203){.}
\put(292,203){.}
\put(298,202){.}
\put(143,235){.}
\put(149,233){.}
\put(155,230){.}
\put(161,228){.}
\put(167,225){.}
\put(173,223){.}
\put(179,221){.}
\put(185,219){.}
\put(191,217){.}
\put(197,216){.}
\put(203,214){.}
\put(209,212){.}
\put(215,211){.}
\put(221,210){.}
\put(227,208){.}
\put(233,207){.}
\put(239,206){.}
\put(245,205){.}
\put(251,204){.}
\put(257,203){.}
\put(263,203){.}
\put(269,202){.}
\put(275,201){.}
\put(281,201){.}
\put(287,200){.}
\put(293,199){.}
\put(139,233){.}
\put(145,230){.}
\put(151,228){.}
\put(157,225){.}
\put(163,223){.}
\put(169,221){.}
\put(175,219){.}
\put(181,217){.}
\put(187,215){.}
\put(193,213){.}
\put(199,211){.}
\put(205,210){.}
\put(211,208){.}
\put(217,207){.}
\put(223,206){.}
\put(229,205){.}
\put(235,203){.}
\put(241,202){.}
\put(247,201){.}
\put(253,201){.}
\put(259,200){.}
\put(265,199){.}
\put(271,198){.}
\put(277,197){.}
\put(283,197){.}
\put(289,196){.}
\put(135,231){.}
\put(141,228){.}
\put(147,225){.}
\put(153,223){.}
\put(159,220){.}
\put(165,218){.}
\put(171,216){.}
\put(177,214){.}
\put(183,212){.}
\put(189,210){.}
\put(195,209){.}
\put(201,207){.}
\put(207,205){.}
\put(213,204){.}
\put(219,203){.}
\put(225,202){.}
\put(231,200){.}
\put(237,199){.}
\put(243,198){.}
\put(249,197){.}
\put(255,196){.}
\put(261,196){.}
\put(267,195){.}
\put(273,194){.}
\put(279,193){.}
\put(285,193){.}
\put(131,228){.}
\put(137,225){.}
\put(143,222){.}
\put(149,220){.}
\put(155,218){.}
\put(161,215){.}
\put(167,213){.}
\put(173,211){.}
\put(179,209){.}
\put(185,207){.}
\put(191,205){.}
\put(197,204){.}
\put(203,202){.}
\put(209,201){.}
\put(215,200){.}
\put(221,198){.}
\put(227,197){.}
\put(233,196){.}
\put(239,195){.}
\put(245,194){.}
\put(251,193){.}
\put(257,192){.}
\put(263,191){.}
\put(269,191){.}
\put(275,190){.}
\put(281,189){.}
\put(126,225){.}
\put(132,222){.}
\put(138,219){.}
\put(144,217){.}
\put(150,214){.}
\put(156,212){.}
\put(162,210){.}
\put(168,208){.}
\put(174,206){.}
\put(180,204){.}
\put(186,202){.}
\put(192,201){.}
\put(198,199){.}
\put(204,198){.}
\put(210,196){.}
\put(216,195){.}
\put(222,194){.}
\put(228,193){.}
\put(234,192){.}
\put(240,191){.}
\put(246,190){.}
\put(252,189){.}
\put(258,188){.}
\put(264,187){.}
\put(270,186){.}
\put(276,186){.}
\put(122,222){.}
\put(128,219){.}
\put(134,216){.}
\put(140,214){.}
\put(146,211){.}
\put(152,209){.}
\put(158,207){.}
\put(164,205){.}
\put(170,203){.}
\put(176,201){.}
\put(182,199){.}
\put(188,197){.}
\put(194,196){.}
\put(200,194){.}
\put(206,193){.}
\put(212,191){.}
\put(218,190){.}
\put(224,189){.}
\put(230,188){.}
\put(236,187){.}
\put(242,186){.}
\put(248,185){.}
\put(254,184){.}
\put(260,183){.}
\put(266,183){.}
\put(272,182){.}
\put(118,218){.}
\put(124,215){.}
\put(130,213){.}
\put(136,210){.}
\put(142,208){.}
\put(148,205){.}
\put(154,203){.}
\put(160,201){.}
\put(166,199){.}
\put(172,197){.}
\put(178,195){.}
\put(184,194){.}
\put(190,192){.}
\put(196,191){.}
\put(202,189){.}
\put(208,188){.}
\put(214,187){.}
\put(220,185){.}
\put(226,184){.}
\put(232,183){.}
\put(238,182){.}
\put(244,181){.}
\put(250,181){.}
\put(256,180){.}
\put(262,179){.}
\put(268,178){.}
\put(114,215){.}
\put(120,212){.}
\put(126,209){.}
\put(132,207){.}
\put(138,204){.}
\put(144,202){.}
\put(150,200){.}
\put(156,198){.}
\put(162,196){.}
\put(168,194){.}
\put(174,192){.}
\put(180,190){.}
\put(186,189){.}
\put(192,187){.}
\put(198,186){.}
\put(204,184){.}
\put(210,183){.}
\put(216,182){.}
\put(222,181){.}
\put(228,180){.}
\put(234,179){.}
\put(240,178){.}
\put(246,177){.}
\put(252,176){.}
\put(258,175){.}
\put(264,175){.}
\put(109,211){.}
\put(115,208){.}
\put(121,206){.}
\put(127,203){.}
\put(133,201){.}
\put(139,198){.}
\put(145,196){.}
\put(151,194){.}
\put(157,192){.}
\put(163,190){.}
\put(169,188){.}
\put(175,186){.}
\put(181,185){.}
\put(187,183){.}
\put(193,182){.}
\put(199,181){.}
\put(205,179){.}
\put(211,178){.}
\put(217,177){.}
\put(223,176){.}
\put(229,175){.}
\put(235,174){.}
\put(241,173){.}
\put(247,172){.}
\put(253,171){.}
\put(259,171){.}
\put(105,208){.}
\put(111,205){.}
\put(117,202){.}
\put(123,200){.}
\put(129,197){.}
\put(135,195){.}
\put(141,193){.}
\put(147,190){.}
\put(153,188){.}
\put(159,186){.}
\put(165,185){.}
\put(171,183){.}
\put(177,181){.}
\put(183,180){.}
\put(189,178){.}
\put(195,177){.}
\put(201,176){.}
\put(207,174){.}
\put(213,173){.}
\put(219,172){.}
\put(225,171){.}
\put(231,170){.}
\put(237,169){.}
\put(243,168){.}
\put(249,168){.}
\put(255,167){.}
\put(101,204){.}
\put(107,201){.}
\put(113,199){.}
\put(119,196){.}
\put(125,193){.}
\put(131,191){.}
\put(137,189){.}
\put(143,187){.}
\put(149,185){.}
\put(155,183){.}
\put(161,181){.}
\put(167,179){.}
\put(173,177){.}
\put(179,176){.}
\put(185,174){.}
\put(191,173){.}
\put(197,172){.}
\put(203,171){.}
\put(209,169){.}
\put(215,168){.}
\put(221,167){.}
\put(227,166){.}
\put(233,165){.}
\put(239,165){.}
\put(245,164){.}
\put(251,163){.}
\put(97,200){.}
\put(103,198){.}
\put(109,195){.}
\put(115,192){.}
\put(121,190){.}
\put(127,187){.}
\put(133,185){.}
\put(139,183){.}
\put(145,181){.}
\put(151,179){.}
\put(157,177){.}
\put(163,175){.}
\put(169,174){.}
\put(175,172){.}
\put(181,171){.}
\put(187,169){.}
\put(193,168){.}
\put(199,167){.}
\put(205,166){.}
\put(211,164){.}
\put(217,163){.}
\put(223,163){.}
\put(229,162){.}
\put(235,161){.}
\put(241,160){.}
\put(247,159){.}
\put(92,197){.}
\put(98,194){.}
\put(104,191){.}
\put(110,189){.}
\put(116,186){.}
\put(122,184){.}
\put(128,181){.}
\put(134,179){.}
\put(140,177){.}
\put(146,175){.}
\put(152,173){.}
\put(158,172){.}
\put(164,170){.}
\put(170,168){.}
\put(176,167){.}
\put(182,165){.}
\put(188,164){.}
\put(194,163){.}
\put(200,162){.}
\put(206,161){.}
\put(212,160){.}
\put(218,159){.}
\put(224,158){.}
\put(230,157){.}
\put(236,156){.}
\put(242,155){.}
\put(88,193){.}
\put(94,190){.}
\put(100,187){.}
\put(106,185){.}
\put(112,182){.}
\put(118,180){.}
\put(124,178){.}
\put(130,175){.}
\put(136,173){.}
\put(142,171){.}
\put(148,169){.}
\put(154,168){.}
\put(160,166){.}
\put(166,164){.}
\put(172,163){.}
\put(178,162){.}
\put(184,160){.}
\put(190,159){.}
\put(196,158){.}
\put(202,157){.}
\put(208,156){.}
\put(214,155){.}
\put(220,154){.}
\put(226,153){.}
\put(232,152){.}
\put(238,151){.}
\put(84,189){.}
\put(90,186){.}
\put(96,184){.}
\put(102,181){.}
\put(108,178){.}
\put(114,176){.}
\put(120,174){.}
\put(126,172){.}
\put(132,170){.}
\put(138,168){.}
\put(144,166){.}
\put(150,164){.}
\put(156,162){.}
\put(162,161){.}
\put(168,159){.}
\put(174,158){.}
\put(180,156){.}
\put(186,155){.}
\put(192,154){.}
\put(198,153){.}
\put(204,152){.}
\put(210,151){.}
\put(216,150){.}
\put(222,149){.}
\put(228,148){.}
\put(234,147){.}
\put(80,185){.}
\put(86,182){.}
\put(92,180){.}
\put(98,177){.}
\put(104,175){.}
\put(110,172){.}
\put(116,170){.}
\put(122,168){.}
\put(128,166){.}
\put(134,164){.}
\put(140,162){.}
\put(146,160){.}
\put(152,158){.}
\put(158,157){.}
\put(164,155){.}
\put(170,154){.}
\put(176,153){.}
\put(182,151){.}
\put(188,150){.}
\put(194,149){.}
\put(200,148){.}
\put(206,147){.}
\put(212,146){.}
\put(218,145){.}
\put(224,144){.}
\put(230,143){.}
\put(75,181){.}
\put(81,179){.}
\put(87,176){.}
\put(93,173){.}
\put(99,171){.}
\put(105,168){.}
\put(111,166){.}
\put(117,164){.}
\put(123,162){.}
\put(129,160){.}
\put(135,158){.}
\put(141,156){.}
\put(147,154){.}
\put(153,153){.}
\put(159,151){.}
\put(165,150){.}
\put(171,149){.}
\put(177,147){.}
\put(183,146){.}
\put(189,145){.}
\put(195,144){.}
\put(201,143){.}
\put(207,142){.}
\put(213,141){.}
\put(219,140){.}
\put(225,139){.}
\put(71,177){.}
\put(77,175){.}
\put(83,172){.}
\put(89,169){.}
\put(95,167){.}
\put(101,165){.}
\put(107,162){.}
\put(113,160){.}
\put(119,158){.}
\put(125,156){.}
\put(131,154){.}
\put(137,152){.}
\put(143,151){.}
\put(149,149){.}
\put(155,147){.}
\put(161,146){.}
\put(167,145){.}
\put(173,143){.}
\put(179,142){.}
\put(185,141){.}
\put(191,140){.}
\put(197,139){.}
\put(203,138){.}
\put(209,137){.}
\put(215,136){.}
\put(221,135){.}
\put(67,174){.}
\put(73,171){.}
\put(79,168){.}
\put(85,166){.}
\put(91,163){.}
\put(97,161){.}
\put(103,158){.}
\put(109,156){.}
\put(115,154){.}
\put(121,152){.}
\put(127,150){.}
\put(133,148){.}
\put(139,147){.}
\put(145,145){.}
\put(151,144){.}
\put(157,142){.}
\put(163,141){.}
\put(169,139){.}
\put(175,138){.}
\put(181,137){.}
\put(187,136){.}
\put(193,135){.}
\put(199,134){.}
\put(205,133){.}
\put(211,132){.}
\put(217,131){.}
\put(63,170){.}
\put(69,167){.}
\put(75,164){.}
\put(81,162){.}
\put(87,159){.}
\put(93,157){.}
\put(99,154){.}
\put(105,152){.}
\put(111,150){.}
\put(117,148){.}
\put(123,146){.}
\put(129,144){.}
\put(135,143){.}
\put(141,141){.}
\put(147,140){.}
\put(153,138){.}
\put(159,137){.}
\put(165,135){.}
\put(171,134){.}
\put(177,133){.}
\put(183,132){.}
\put(189,131){.}
\put(195,130){.}
\put(201,129){.}
\put(207,128){.}
\put(213,127){.}
\put(58,166){.}
\put(64,163){.}
\put(70,160){.}
\put(76,158){.}
\put(82,155){.}
\put(88,153){.}
\put(94,150){.}
\put(100,148){.}
\put(106,146){.}
\put(112,144){.}
\put(118,142){.}
\put(124,140){.}
\put(130,139){.}
\put(136,137){.}
\put(142,136){.}
\put(148,134){.}
\put(154,133){.}
\put(160,132){.}
\put(166,130){.}
\put(172,129){.}
\put(178,128){.}
\put(184,127){.}
\put(190,126){.}
\put(196,125){.}
\put(202,124){.}
\put(208,123){.}
\put(54,162){.}
\put(60,159){.}
\put(66,156){.}
\put(72,154){.}
\put(78,151){.}
\put(84,149){.}
\put(90,147){.}
\put(96,144){.}
\put(102,142){.}
\put(108,140){.}
\put(114,138){.}
\put(120,137){.}
\put(126,135){.}
\put(132,133){.}
\put(138,132){.}
\put(144,130){.}
\put(150,129){.}
\put(156,128){.}
\put(162,126){.}
\put(168,125){.}
\put(174,124){.}
\put(180,123){.}
\put(186,122){.}
\put(192,121){.}
\put(198,120){.}
\put(204,119){.}
\put(50,158){.}
\put(56,155){.}
\put(62,152){.}
\put(68,150){.}
\put(74,147){.}
\put(80,145){.}
\put(86,143){.}
\put(92,140){.}
\put(98,138){.}
\put(104,136){.}
\put(110,134){.}
\put(116,133){.}
\put(122,131){.}
\put(128,129){.}
\put(134,128){.}
\put(140,126){.}
\put(146,125){.}
\put(152,124){.}
\put(158,122){.}
\put(164,121){.}
\put(170,120){.}
\put(176,119){.}
\put(182,118){.}
\put(188,117){.}
\put(194,116){.}
\put(200,115){.}
\put(46,154){.}
\put(52,151){.}
\put(58,148){.}
\put(64,146){.}
\put(70,143){.}
\put(76,141){.}
\put(82,139){.}
\put(88,136){.}
\put(94,134){.}
\put(100,132){.}
\put(106,130){.}
\put(112,129){.}
\put(118,127){.}
\put(124,125){.}
\put(130,124){.}
\put(136,122){.}
\put(142,121){.}
\put(148,120){.}
\put(154,118){.}
\put(160,117){.}
\put(166,116){.}
\put(172,115){.}
\put(178,114){.}
\put(184,113){.}
\put(190,112){.}
\put(196,111){.}
\put(150,200){\vector(0,1){100}}
\put(150,200){\vector(1,0){170}}
\put(150,200){\vector(-1,-1){120}}
\put(140,300){$r_1$}
\put(335,205){$b$}
\put(20,70){$l$}
\put(40,90){\line(1,0){156}}
\put(40,90){\line(0,1){64}}
\put(20,90){5.2}
\put(306,200){\line(-1,-1){110}}
\put(306,200){\line(0,1){8}}
\put(306,215){5.2}
\put(196,90){\line(0,1){22}}
\put(125,245){0.3}
\end{picture}

{ \it

\begin{center}
Fig.1
\end{center}

 The relative error of $N_A$ drawn as the function of $l$ and $b$.
Parameter $l$ goes from $0.2$ up to $5.2$ with a step $0.2$.
Parameter $b$ goes from $0.2$ up to $5.2$ with a step $0.2$.

One can see the essential negative slope when $b$ increases and the slight
positive slope when $l$ increases.

}

\pagebreak

\begin{picture}(350,350)
\put(0,0){\line(0,1){350}}
\put(0,0){\line(1,0){350}}
\put(350,0){\line(0,1){350}}
\put(0,350){\line(1,0){350}}
\put(145,209){.}
\put(151,207){.}
\put(157,206){.}
\put(163,204){.}
\put(169,203){.}
\put(175,202){.}
\put(181,201){.}
\put(187,200){.}
\put(193,199){.}
\put(199,198){.}
\put(205,198){.}
\put(211,197){.}
\put(217,197){.}
\put(223,196){.}
\put(229,196){.}
\put(235,195){.}
\put(241,195){.}
\put(247,194){.}
\put(253,194){.}
\put(259,194){.}
\put(265,193){.}
\put(271,193){.}
\put(277,193){.}
\put(283,192){.}
\put(289,192){.}
\put(295,192){.}
\put(135,202){.}
\put(141,200){.}
\put(147,199){.}
\put(153,197){.}
\put(159,196){.}
\put(165,194){.}
\put(171,193){.}
\put(177,192){.}
\put(183,191){.}
\put(189,190){.}
\put(195,189){.}
\put(201,189){.}
\put(207,188){.}
\put(213,187){.}
\put(219,187){.}
\put(225,186){.}
\put(231,186){.}
\put(237,185){.}
\put(243,185){.}
\put(249,184){.}
\put(255,184){.}
\put(261,184){.}
\put(267,183){.}
\put(273,183){.}
\put(279,183){.}
\put(285,182){.}
\put(124,194){.}
\put(130,192){.}
\put(136,190){.}
\put(142,189){.}
\put(148,187){.}
\put(154,186){.}
\put(160,184){.}
\put(166,183){.}
\put(172,182){.}
\put(178,181){.}
\put(184,180){.}
\put(190,179){.}
\put(196,179){.}
\put(202,178){.}
\put(208,177){.}
\put(214,177){.}
\put(220,176){.}
\put(226,175){.}
\put(232,175){.}
\put(238,175){.}
\put(244,174){.}
\put(250,174){.}
\put(256,173){.}
\put(262,173){.}
\put(268,173){.}
\put(274,172){.}
\put(114,185){.}
\put(120,183){.}
\put(126,181){.}
\put(132,180){.}
\put(138,178){.}
\put(144,176){.}
\put(150,175){.}
\put(156,174){.}
\put(162,173){.}
\put(168,172){.}
\put(174,171){.}
\put(180,170){.}
\put(186,169){.}
\put(192,168){.}
\put(198,167){.}
\put(204,167){.}
\put(210,166){.}
\put(216,166){.}
\put(222,165){.}
\put(228,165){.}
\put(234,164){.}
\put(240,164){.}
\put(246,163){.}
\put(252,163){.}
\put(258,163){.}
\put(264,162){.}
\put(103,176){.}
\put(109,174){.}
\put(115,172){.}
\put(121,170){.}
\put(127,169){.}
\put(133,167){.}
\put(139,166){.}
\put(145,164){.}
\put(151,163){.}
\put(157,162){.}
\put(163,161){.}
\put(169,160){.}
\put(175,159){.}
\put(181,158){.}
\put(187,158){.}
\put(193,157){.}
\put(199,156){.}
\put(205,156){.}
\put(211,155){.}
\put(217,155){.}
\put(223,154){.}
\put(229,154){.}
\put(235,153){.}
\put(241,153){.}
\put(247,152){.}
\put(253,152){.}
\put(92,167){.}
\put(98,165){.}
\put(104,163){.}
\put(110,161){.}
\put(116,159){.}
\put(122,157){.}
\put(128,156){.}
\put(134,155){.}
\put(140,153){.}
\put(146,152){.}
\put(152,151){.}
\put(158,150){.}
\put(164,149){.}
\put(170,148){.}
\put(176,148){.}
\put(182,147){.}
\put(188,146){.}
\put(194,146){.}
\put(200,145){.}
\put(206,144){.}
\put(212,144){.}
\put(218,143){.}
\put(224,143){.}
\put(230,143){.}
\put(236,142){.}
\put(242,142){.}
\put(82,157){.}
\put(88,155){.}
\put(94,153){.}
\put(100,151){.}
\put(106,149){.}
\put(112,148){.}
\put(118,146){.}
\put(124,145){.}
\put(130,143){.}
\put(136,142){.}
\put(142,141){.}
\put(148,140){.}
\put(154,139){.}
\put(160,138){.}
\put(166,137){.}
\put(172,137){.}
\put(178,136){.}
\put(184,135){.}
\put(190,135){.}
\put(196,134){.}
\put(202,134){.}
\put(208,133){.}
\put(214,133){.}
\put(220,132){.}
\put(226,132){.}
\put(232,132){.}
\put(71,148){.}
\put(77,145){.}
\put(83,143){.}
\put(89,141){.}
\put(95,140){.}
\put(101,138){.}
\put(107,136){.}
\put(113,135){.}
\put(119,134){.}
\put(125,132){.}
\put(131,131){.}
\put(137,130){.}
\put(143,129){.}
\put(149,128){.}
\put(155,127){.}
\put(161,127){.}
\put(167,126){.}
\put(173,125){.}
\put(179,125){.}
\put(185,124){.}
\put(191,123){.}
\put(197,123){.}
\put(203,123){.}
\put(209,122){.}
\put(215,122){.}
\put(221,121){.}
\put(61,138){.}
\put(67,136){.}
\put(73,134){.}
\put(79,132){.}
\put(85,130){.}
\put(91,128){.}
\put(97,126){.}
\put(103,125){.}
\put(109,124){.}
\put(115,122){.}
\put(121,121){.}
\put(127,120){.}
\put(133,119){.}
\put(139,118){.}
\put(145,117){.}
\put(151,116){.}
\put(157,116){.}
\put(163,115){.}
\put(169,114){.}
\put(175,114){.}
\put(181,113){.}
\put(187,113){.}
\put(193,112){.}
\put(199,112){.}
\put(205,111){.}
\put(211,111){.}
\put(50,128){.}
\put(56,126){.}
\put(62,124){.}
\put(68,122){.}
\put(74,120){.}
\put(80,118){.}
\put(86,116){.}
\put(92,115){.}
\put(98,113){.}
\put(104,112){.}
\put(110,111){.}
\put(116,110){.}
\put(122,109){.}
\put(128,108){.}
\put(134,107){.}
\put(140,106){.}
\put(146,105){.}
\put(152,105){.}
\put(158,104){.}
\put(164,103){.}
\put(170,103){.}
\put(176,102){.}
\put(182,102){.}
\put(188,101){.}
\put(194,101){.}
\put(200,101){.}
\put(39,118){.}
\put(45,116){.}
\put(51,114){.}
\put(57,112){.}
\put(63,110){.}
\put(69,108){.}
\put(75,106){.}
\put(81,105){.}
\put(87,103){.}
\put(93,102){.}
\put(99,101){.}
\put(105,100){.}
\put(111,99){.}
\put(117,98){.}
\put(123,97){.}
\put(129,96){.}
\put(135,95){.}
\put(141,94){.}
\put(147,94){.}
\put(153,93){.}
\put(159,93){.}
\put(165,92){.}
\put(171,92){.}
\put(177,91){.}
\put(183,91){.}
\put(189,90){.}
\put(150,200){\vector(0,1){100}}
\put(150,200){\vector(1,0){170}}
\put(150,200){\vector(-1,-1){120}}
\put(140,300){$r_1$}
\put(335,205){$b$}
\put(20,70){$l$}
\put(33,83){\line(1,0){156}}
\put(33,83){\line(0,1){35}}
\put(13,83){0.1}
\put(306,200){\line(-1,-1){117}}
\put(306,200){\line(0,1){3}}
\put(306,215){5.2}
\put(189,83){\line(0,1){7}}
\put(125,245){0.3}
\end{picture}

\begin{center}
Fig.2
\end{center}

{ \it
 The relative error of $N_A$ drawn as the function of $l$ and $b$.
Parameter $l$ goes from $0.01$ up to $0.11$ with a step $0.01$.
Parameter $b$ goes from $0.2$ up to $5.2$ with a step $0.2$.

One can see the essential negative slope when $b$ increases and the slight
positive slope when $l$ increases. The qualitative character is absolutely
the same as in fig. 1.

}

\end{document}